\documentclass[prd,twocolumn,floats,floatfix,showpacs,nofootinbib]{revtex4}
\usepackage{graphicx}
\usepackage{dcolumn}
\usepackage{bm}
\usepackage{graphics}
\usepackage{slashed}
\usepackage{amssymb}
\usepackage{natbib}
\usepackage{amsmath}
\usepackage{url}
\usepackage{color}
\newcommand{\bea}{\begin{eqnarray}}
\newcommand{\ena}{\end{eqnarray}}
\newcommand{\bean}{\begin{eqnarray*}}
\newcommand{\enan}{\end{eqnarray*}}

\begin{document}

\title{The cosmological origin of the Nambu-Jona-Lasinio model}

\author{M. Novello\footnote{M. Novello is Cesare Lattes ICRANet Professor}} \email{novello@cbpf.br}

\affiliation{Instituto de Cosmologia Relatividade Astrofisica ICRA -
CBPF\\ Rua Dr. Xavier Sigaud, 150, CEP 22290-180, Rio de Janeiro,
Brazil}

\date{\today}

\begin{abstract}
Recently a mechanism to generate mass from gravitational
interaction, based on Mach principle, according to which the inertia
of a body is a property of matter as well as of the background
provided by the rest-of-the-universe was presented in \cite{novello}
\cite{novello2}. In these papers  such an idea was realized for
scalar and spinor fields treating the rest-of-the-universe in its
vacuum state. In the present paper, using an extended version of
Mach principle, the same strategy will be applied to show how the
 Heisenberg-Nambu-Jona-Lasinio non-linear equation for fermions $\Psi$
arises as a consequence of the gravitational interaction of $ \Psi$
with the rest-of-the-universe.
\end{abstract}

\maketitle

\section{Introduction}

Although the theory of General Relativity may be understood as
completely independent from the Machian idea that inertia of a body
$\mathbb{A}$ is related to the global distribution of energy of all
particles existing in the universe, we must recognize its historical
value in the making the ideology that enabled Einstein to start his
journey toward the construction of a theory of gravitation
\cite{vizgin} \cite{dicke}.

During the 20th century, the idea of associating the dependence of
local characteristics of matter with the global state of the
universe came up now and then but without producing any reliable
mechanism that could support such proposal.

Even the concept of mass -- that pervades all gravitational
processes -- did not find a realization of such dependence on global
structure of the universe. On the contrary, the most efficient
mechanism and one that has performed an important role in the field
of microphysics came from elsewhere, namely the attempt to unify
forces of a non-gravitational character, such as long-range
electrodynamics with decaying phenomena described by weak
interaction. Indeed, the Higgs model produced an efficient scenario
for generating mass to the vector bosons \cite{halzen} that goes in
the opposite direction of the proposal of Mach. This mechanism
starts with the transformation of a global symmetry into a local one
and the corresponding presence of vector gauge fields. Then, a
particular form of the dynamics represented by $ L_{int}(\varphi)$
of self-interaction of an associated scalar field in its fundamental
state represented by an energy-momentum tensor given by $ T_{\mu\nu}
= L_{int}(\varphi_{0}) \, g_{\mu\nu}$ appears as the vehicle which
provides mass to the gauge fields.

Recently a new mechanism for generation of mass that is a
realization of Mach\rq s idea was proposed \cite{novello}. The
strategy used in that paper is the following. We start with a theory
of a massless scalar field in \cite{novello} (and for fermions in
\cite{novello2}) coupled non-minimally to gravity through an
interacting Lagrangian of the form $ V(\varphi) \, R. $ The
distribution of the vacuum energy of the rest-of-the-universe is
represented by a cosmological term $\Lambda.$ The effect of $
\Lambda$ by the intermediary of the dynamics of the metric of
space-time in the realm of General Relativity is precisely to give
mass to the field. Although this mass depends on the cosmological
constant, its value cannot be obtained a priori.

In the present paper we will apply this strategy in order to
generalize Mach \rq s idea, and following Dirac, Hoyle and others,
to produce a mechanism that can transform the vague idea according
to which local properties may depend on the universe\rq s global
characteristics, into an efficient process.

The first question we have to face concerns the choice of the
elementary process. There have been many discussions in the
scientific literature in the last decades related to the cosmic
dependence of the fundamental interactions. We will do not analyze
any of these here. On the contrary, we will concentrate on a
specific self-interaction of an elementary field and show that its
correspondent dynamics is a consequence of a dynamical cosmological
process. That is, to show that dynamics of elementary fields in the
realm of microphysics, may depend on the global structure of the
universe.

 There is no better way than start our analysis with the
fundamental theory proposed by Nambu and Jona-Lasinio concerning a
dynamical model of elementary particles \cite{nambu}. Since the
original paper until to-day hundreds of papers devoted to the NJL
model were published \cite{volkov}. For our purpose here it is
enough to analyze the nonlinear equation of motion that they used in
their original paper as the basis of their theory which is

$$ i\gamma^{\mu} \nabla_{\mu} \, \Psi  - 2s ( A + i \, B \, \gamma^{5} )\Psi =
0 $$ This equation, as remarked by these authors, was proposed
earlier by Heisenberg \cite{heisenberg} although in a quite
different context. We will not enter in the analysis of the theory
that follows from this dynamics. Our question here is just this: is
it possible to produce a model such that HNJL
(Heisenberg-Nambu-Jona-Lasinio) equation for
 spinor field becomes a consequence of the gravitational interaction of a free massless Dirac
 field with the rest-of-the-universe? I claim that the answer is
 yes. This article was written to prove this.

\section{The cosmological influence on the microphysical world: the case
of chiral-invariant Heisenberg-Nambu-Jona-Lasinio dynamics}

We take Mach\rq s principle as the statement according to which the
inertial properties of a body $\mathbb{A }$ are determined by the
energy-momentum throughout all space. The description of such
universal state that takes into account the whole contribution of
the rest-of-the-universe onto $\mathbb{A }$ is the most homogeneous
 one and is related to what Einstein attributed to the cosmological
constant or, in modern language, the vacuum of all remaining bodies. We
follow a similar procedure here and will consider that the Extended Mach Principle (EMP)
means that the influence of the rest-of-the-universe on any elementary
field in microphysics can be described through the identification of
the energy-momentum distribution of all remaining bodies under the form
$$ T^{U}_{\mu\nu} = \Lambda
\, g_{\mu\nu}
$$

\subsection{Non minimal coupling with gravity}

In the framework of General Relativity we set the dynamics of a
fermion field $\Psi$ coupled non-minimally with gravity to be given
by the Lagrangian (we are using units were $\hbar = c = 1)$

\begin{equation}
L = L_{D} + \frac{1}{\kappa} \, R +  V(X) \, R - \frac{1}{\kappa}
 \, \Lambda + L_{CT}
\label{401}
\end{equation}
where
\begin{equation}
L_{D} \equiv \frac{i}{2} \bar{\Psi} \gamma^{\mu} \nabla_{\mu} \Psi -
\frac{i}{2} \nabla_{\mu} \bar{\Psi} \gamma^{\mu} \Psi
 \label{302}
\end{equation}
The non-minimal coupling of the spinor field with gravity is
contained in the term $ V(X)$ and depends on the scalar $ X$ defined
by  $$ X = A^{2} + B^{2}$$ where $ A = \bar{\Psi} \, \Psi $ and $B =
i \bar{\Psi} \, \gamma^{5} \, \Psi.$ We note that we can write, in
an equivalent way,  $$X = J_{\mu} \, J^{\mu} $$ where $ J^{\mu} =
\bar{\Psi} \gamma^{\mu}  \Psi.$ This quantity $ X $ is chiral
invariant, once it is invariant under the map
$$ \Psi' = \gamma^{5} \, \Psi.$$ Indeed, from this $\gamma^{5}$ transformation,
it follows

$$ A' = - \, A, \,\,   B' = - \, B; \,\,  then,  X' = X.$$

The case in which the theory breaks chiral invariance the potential
$ V $ depends only on the invariant $A$ -- and it is the road to the
appearance of a mass \cite{novello2}. Here we start from the
beginning with a chiral invariant theory. For the time being the
dependence of $ V $ on $ X $ is not fixed. We have added $L_{CT}$ to
counter-balance the terms of the form $
\partial_{\lambda} X \, \partial^{\lambda} X $ and $ \Box X $
that appear due to the gravitational interaction. The most general
form of this counter-term is
\begin{equation}L_{CT} = H(X) \, \partial_{\mu} X \, \partial^{\mu} X
\label{403}\end{equation} We shall see that $ H $ depends on $ V$
and if we set $ V = 0 $ then $ H $ vanishes. This dynamics
represents a massless spinor field coupled non-minimally with
gravity. The cosmological constant is added by the reasons presented
above and as we shall see it represents the influence of the
rest-of-the-universe on $\Psi.$

Independent variation of $\Psi$ and $g_{\mu\nu}$ yields
\begin{equation}
 i\gamma^{\mu} \nabla_{\mu} \, \Psi  +  \Omega   \, ( A + i \, B \, \gamma^{5} )\Psi = 0 \label{404}
\end{equation}
where
$$  \Omega \equiv 2 R \, V'  -
 2 H' \, \partial_{\mu} X \, \partial^{\mu} X  - 4 H \Box X $$

\begin{equation}
\alpha_{0} \, ( R_{\mu\nu} - \frac{1}{2} \, R \, g_{\mu\nu} ) = -
T_{\mu\nu}
 \label{405}
\end{equation}
where we set  $ \alpha_{0} \equiv 2 /\kappa $ and $ V' \equiv
\partial V / \partial X.$ The energy-momentum tensor defined by

 $$T_{\mu\nu} = \frac{2}{\sqrt{- g}} \, \frac{\delta ( \sqrt{-g} \,
 L)}{\delta g^{\mu\nu}} $$
is given by
\begin{eqnarray}
T_{\mu\nu} &=& \frac{i}{4} \, \bar{\Psi} \gamma_{(\mu} \nabla_{\nu)}
\Psi - \frac{i}{4} \nabla_{(\mu} \bar{\Psi} \gamma_{\mu)} \Psi
\nonumber \\
&+& 2 V ( R_{\mu\nu} - \frac{1}{2} \, R \, g_{\mu\nu} ) + 2
\nabla_{\mu} \nabla_{\nu} V - 2 \Box V g_{\mu\nu} \nonumber \\
&+& 2 H \, \partial_{\mu} X \, \partial_{\nu} X - H \,
\partial_{\lambda} X \, \partial^{\lambda} X \, g_{\mu\nu} +
\frac{\alpha_{0}}{2} \, \Lambda \, g_{\mu\nu} \label{406}
 \end{eqnarray}

Taking the trace of equation (\ref{405}), after some simplification
and using
\begin{equation}
 \Box V = V' \, \Box X + V'' \, \partial_{\mu} X \, \partial^{\mu}
X
\end{equation}
 it follows
\begin{eqnarray}
( \alpha_{0} + 2 V + 2 \, V' \, X) \, R &=& (4 H X - 6 V') \Box X
\nonumber
\\ &+& ( 2 H'\, X - 6 V'' - 2 H) \, \partial_{\alpha} X \,
\partial^{\alpha} X \nonumber \\
&+& 2 \, \alpha_{0} \, \Lambda
\end{eqnarray}
Then

\begin{eqnarray}
\Omega &=&  \left( \mathbb{M} \, \Box X + \mathbb{N} \,
\partial_{\mu} X \,
\partial^{\mu} X \right) \nonumber
\\
&+& \frac{4 \, \alpha_{0} \, \Lambda \, V'}{\alpha_{0} + 2 V + 2 \,
V' \, X}
\end{eqnarray}
where
$$ \mathbb{M} = \frac{2 V' ( 4 H X - 6 V')}{ \alpha_{0} + 2 V + 2 \, V' \,
X }  - 4 \, H $$

$$ \mathbb{N} = \frac{2 V' \, ( 2 \, X \, H' -  6 V'' - 2 \, H)}{\alpha_{0} + 2 V + 2 \, V' \, X}  - 2 \, H'
$$

Defining $ \Delta \equiv \alpha_{0} + 2 V + 2 \, V' \, X $ we
re-write $\mathbb{M}$ and $\mathbb{N}$ as
$$  \mathbb{M} = - \, \frac{4}{\Delta} \, \left( 3 \, V'^{2} + H \,(\alpha_{0} + 2 V ) \right) $$
$$  \mathbb{N} = - \, \frac{2}{\Delta} \, \left( 3 \, V'^{2} + H \,(\alpha_{0} + 2 V ) \right)' $$

Inserting this result on the equation (\ref{404}) yields
\begin{equation}
 i\gamma^{\mu} \nabla_{\mu} \, \Psi  +
\left( \mathbb{M} \, \Box X  + \mathbb{N} \,
\partial_{\lambda} X \, \partial^{\lambda} X \right) \,
\Psi + \mathbb{Z} \, ( A + i \, B \, \gamma^{5} )\Psi = 0
\label{407}
\end{equation}
where
$$  \mathbb{Z} = \frac{ 4 \,\alpha_{0} \, \Lambda \, V'}{\alpha_{0} + 2  V + 2 \, V' \, X}
  $$

At this stage it is worth to select among all possible candidates of
$ V$ and $ H $ particular ones that makes the factor on the gradient
and on $ \Box $ of the field to disappear from equation (\ref{407}).

The simplest way is to set  $ \mathbb{M} = \mathbb{N} = 0,$  which
is satisfied if
$$H = - \, \frac{3 \, V'^{2}}{\alpha_{0} + 2 V} $$

Imposing that  $ \mathbb{Z}$ must reduce to a constant we obtain
\begin{equation}
V = \frac{1}{\kappa} \, \left[ \frac{1}{ 1 + \beta \, X} - 1
\right]. \label{408}
\end{equation}
As a consequence of this,
\begin{equation}
H = - \, \frac{3 \, \beta^{2}}{2 \kappa}  \, \frac{1}{(1 +  \beta \,
X)^{3}} \label{409}
\end{equation}
where $ \beta $ is a constant. Note that $ V $ vanishes when $ \beta
= 0,$ and as a consequence, $ H $ vanishes too.

The equation for the spinor becomes

\begin{equation}
 i\gamma^{\mu} \nabla_{\mu} \, \Psi  - 2s ( A + i \, B \, \gamma^{5} )\Psi = 0  \label{410}
\end{equation}
where \begin{equation}
 s = \frac{2 \, \beta \, \Lambda}{\kappa (\hbar \, c)}.
 \label{20}
 \end{equation}

Thus as a result of the above process the field satisfies
Heisenberg-Nambu-Jona-Lasinio equation of motion. This is possible
due to the influence of the rest-of-the-Universe on $ \Psi.$  If $
\Lambda $ vanishes then the constant of the self-interaction of $
\Psi$ vanishes. This is precisely what we envisaged to obtain: the
net effect of the non-minimal coupling of gravity with the spinor
field corresponds to a specific self-interaction. The HNJL equation
of motion appears only if we take into account the existence of all
remaining bodies in the universe --- represented by the cosmological
constant --- in the state in which all existing matter is on the
corresponding vacuum.

The various steps of our mechanism can be synthesized as follows:
\begin{itemize}
\item{The dynamics of a massles spinor field $ \Psi$ is described
by the Lagrangian
$$ L_{D} = \frac{i}{2} \bar{\Psi} \gamma^{\mu} \nabla_{\mu} \Psi -
\frac{i}{2} \nabla_{\mu} \bar{\Psi} \gamma^{\mu} \Psi ; $$}
\item{Gravity is described in General Relativity by the scalar of
curvature $$ L_{E} = R ;$$ }
\item{The field interacts with gravity in a non-minimal way described
by the term
$$ L_{int} = V(X) \, R $$
where $ X = A^{2} + B^{2}$ and $ A = \bar{\Psi} \, \Psi $ and $B = i
\bar{\Psi} \, \gamma^{5} \, \Psi$}
\item{The action of the rest-of-the-universe on the spinor field,
through the gravitational intermediary, is contained in the form of
an additional constant term on the Lagrangian noted as $ \Lambda ;$
}
\item{A counter-term depending on the invariant $ X $
is introduced to kill extra terms coming from gravitational
interaction;}
\item{As a result of this process, after choosing $ V $ and $ H $ the field acquires a self-interaction term
and its equation of motion is precisely
Heisenberg-Nambu-Jona-Lasinio described as
$$ i\gamma^{\mu} \nabla_{\mu} \, \Psi  - 2s \, (A + iB \gamma^{5}) \, \Psi= 0   $$ where
$ s $ is given by equation (\ref{20}) and is zero only if the
cosmological constant vanishes.}
\end{itemize}

It is not hard to envisage others situations in which the above
mechanism can be further applied. Then the question appears:
how far one can extend this argument?  We will come back to this in a future work.

\section{Conclusion}

In this paper we considered the influence of all the material
content of the universe on a fermionic field when this content is in
two possible states: in one case its energy distribution is zero; in
another case it is in a vacuum state represented by the homogeneous
distribution $ T_{\mu\nu} = \Lambda g_{\mu\nu}.$ In the first case
the dynamics of the field is independent of the global properties of
the universe and is described by the massless Dirac equation

$$ i\gamma^{\mu} \nabla_{\mu} \, \Psi  = 0 $$

In the second case, the rest-of-the-universe induces on field  $\Psi$ the
Heisenberg-Nambu-Jona-Lasinio non-linear dynamics

$$  i\gamma^{\mu} \nabla_{\mu} \, \Psi  -
2s \, (A + iB \gamma^{5}) \, \Psi= 0.$$

Thus, the scenario presented in this paper led us to consider the
idea that microphysics can depend strongly on the global structure
of the universe.

\section{acknowledgements}
I would like to thank FINEP, CNPq and FAPERJ for financial support.
I thank Ugo Moschella for conversations on the Nambu-Jona-Lasinio
paper.

\end{document}